\documentclass[aps,prb,twocolumn,groupedaddress,floatfix,showpacs]{revtex4-2}
\usepackage{graphicx}
\usepackage{amsmath}
\usepackage{subcaption}
\usepackage[T1]{fontenc}
\usepackage{color}

\begin{document}

%\preprint{}

\title{Copper plasmonics with excitons}


\author{David Ziemkiewicz}
\email{david.ziemkiewicz@utp.edu.pl}

%\thanks{}
%\author{Karol Karpi\'{n}ski}
\author{Sylwia Zieli\'{n}ska-Raczy\'{n}ska}

 \affiliation{Institute of
Mathematics and Physics, Technical University of Bydgoszcz,
\\ Al. Prof. S. Kaliskiego 7, 85-789 Bydgoszcz, Poland}


\date{\today}

\begin{abstract} 
We investigate the propagation of surface plasmons (SPPs) in a thin layer of copper surrounded by copper oxide Cu$_2$O. It is shown that particularly strong excitons in Cu$_2$O can have considerable impact on plasmon propagation, providing many opportunities for plasmon-exciton and plasmon-plasmon interactions. It is demonstrated that by the use of sufficiently thin metal layer, one can excite the so-called long range plasmons (LRSPPs) which can overcome the inherently high ohmic losses of Cu as compared to usual plasmonic metals such as silver. Analytical results are confirmed by numerical calculations.
\end{abstract}

%\pacs{78.20.-e, 71.35.Cc, 71.36.+c}
%\keywords{}

\maketitle
\section{Introduction}

The surface plasmon-polaritons (SPPs) are bound, nonradiative electromagnetic excitations associated with charge density waves propagating along the metal/dielectric interface. It is well known that surface plasmons can be used to confine the field of electromagnetic wave into subwavelength areas, greatly increasing field intensity and as a result the intensity of electromagnetic field is enhanced. Plasmonic devices offer many significant advantages in solid state device applications; electromagnetic wave bound to a metallic surface can be not only focused, but also easily guided or slowed down. All these properties are very useful for providing a favourable environment for light - exciton interaction. Development of applications where both energy propagation and field confinement are desired requires detailed analysis of the trade-off between the two. This interchange is a direct consequence of a field confinement and plasmon propagation being reciprocal to each other, i.e. propagating SPPs assume localized character in frequency region of surface plasmon while as confinement gets stronger, the electric field penetrates deeper into the metal (surface), leading to increased energy losses. While noble metals are the most popular among plasmonic materials,
copper-based plasmonic devices are a recent, dynamically expanding area of research, with promising applications in nanoantennas \cite{Bohme19}, photovoltaics \cite{Melo18,Huang20}, waveguides \cite{Fedyanin16}.

A nontrivial aspect of design  copper-based devices is the surface protection from oxidation. Various protective layers were proposed, including graphene \cite{Kravets14}. Without the oxide layer, Cu nanostructures exhibit a narrow plasmonic resonance comparable to the one found in silver and gold nanostructures \cite{Chan07}. On the other hand, a controlled layer of oxide with specific thickness might enhance the plasmonic resonance instead of perturbing it \cite{Rodriguez11,Parramon19}. The ongoing research towards reduction of ohmic losses in copper nanostructures leads to geometries that can match the performance of the usual plasmonic materials such as gold and silver \cite{Mkhitaryan21}.

An exciton in a semiconductor is a bound state of a conduction band electron and a valence band hole, which are attracted to each other by the Coulomb interaction; they form an electrically neutral quasiparticle, transferring the energy without  transporting net electric charge.
The ability to couple plasmons with excitons, creating so-called plexcitons \cite{Fofang08,Karademir14} leads to a variety of plasmonic devices that can be tuned on demand \cite{Lee16,Cao18,Goncalves18}. Exciton-plasmon coupling has been demonstrated in the case of Wannier excitons in semiconductors \cite{Khurgin19}, but the use of copper oxide, in particular Cu - Cu$_2$O structures is a novel idea. 
Excitonic states in Cu$_2$O are a solid-state analogue of hydrogen atoms and states with principal quantum number up to  $n=$25 for dipole-allowed P-type envelope wavefunctions, called Rydberg excitons (REs), have been observed \cite{Kazimierczuk}.  Rydberg excitons are becoming one of the most versatile, scalable and tunable platform for quantum computing technologies. The field is developing very rapidly, with first theoretical studies \cite{Poddubny,DZ9} and experiments \cite{Hamid,Konzelmann} exploring nonlinear properties of RE in low-dimensional systems being performed right now. The fabrication techniques are just entering the stage where consistent synthesis of high-quality Cu$_2$O nanostructures becomes possible \cite{Steinhauer,Takahata2018}.  Furthermore, from a more general point of view, Rydberg excitons are one of a kind structures that may offer new insight into fundamental physics; their huge dimension make them an ideal candidate for performing experiments with spatially engineered light fields and micrometer size plasmonic systems. The prominent idea of this paper is to take advantage of their unusual properties coupling these structures with plasmons

The paper is organized as follows: in the next section the basic properties of Rydberg excitons important to the propagation of plasmons are discussed and some general limitations of the system are outlined. Next, the dispersive properties of plasmons excited on a thin metal layer are investigated focusing on the specific case of copper surrounded by copper oxide. Section IV contains calculation results. Finally, the details of employed numerical method are given in the appendix and the last section presents the summary of our results.

\section{Excitons and Rydberg blockade}
An exciton, being an excited electronic state of the crystal is an electron-hole air, which is weakly bound and can  extend over many thousands of lattice unit cells, thus its interaction is screened by static permittivity of the semiconductor. The first observation of higher excited excitons, called Rydberg excitons, first in a cuprous oxide bulk \cite{Kazimierczuk} and then in nanostructures \cite{Hamid} revealed a lot of their  astonishing features, such as extraordinary large dimension scaling as $n^2$, long life-times $\sim n^3$, reaching nanoseconds. Their  extraordinary vulnerability to interactions with external fields is due to huge polarizability scaling as $n^7$.

Another characteristic property of excitons is the so-called Rydberg blockade. For a given  exciton with principal quantum number $n$, another exciton cannot be created in immediate vicinity due to the exciton-exciton interaction that shifts the energy levels. The space where another exciton cannot be created is described by a so-called blockade volume \cite{Kazimierczuk}
\begin{equation}
V_B = 3 \cdot 10^{-7} n^7~\mu m^3.
\end{equation}
The crucial feature is the extremely fast $n^7$ scaling of the blockade; highly excited states very quickly reach the saturation level when the medium is completely filled with excitons. As a result, the light propagating through the medium is not absorbed to create new excitons. This is so-called optical bleaching. On the other hand, the lowest excitonic states can reach considerable density. Specifically, for $n=2$, taking into account a semi-random distribution of blockade volumes as opposed to ideal sphere packing \cite{Thomas2022} one obtains an upper density limit on the order of $10^{17}$ $cm^{-3}$, with some sources reporting even $10^{19}$ $cm^{-3}$ densities \cite{Kavoulakis95}. The issue of maximum possible density is important to the effect of excitonic transitions on surface plasmons; specifically, the propagation properties of SPP are dependent on the Cu$_2$O permittivity $\epsilon = 7.5 + \chi$, where the changes of susceptibility $\chi$ caused by excitons are directly proportional to exciton density. It is beneficial to obtain conditions where these susceptibility changes are not negligible when compared with the ,,static'' part $\epsilon_b=7.5$. This can be only achieved with the lowest excitonic states. Furthermore, another reason to consider only the lowest P-excitonic states ($n$=2,3) is the interaction of excitons with metallic surfaces. As mentioned above, excitons with high principal quantum number are very sensitive to external electric fields, including electrostatic fields at metal-dielectric interface \cite{Kolhoff16}, leading to ionisation of excitons that are separated from the metal surface by a distance comparable to exciton radius. Thus, for example, one can expect that $n=2$ exciton in Cu$_2$O is strongly affected by the metallic environment when located closer than approximately 6 nm from the metal-dielectric interface.

\section{Propagation of surface plasmons}
Surface plasmon is a localized electromagnetic excitation on an interface between two media with exhibit opposite signs of dielectric permittivity at some frequency $\omega$. Typically, the material with negative permittivity $\epsilon_1(\omega)$ is a metal and the other one characterized by $\epsilon_2(\omega)>0$ is a dielectric. In such a system, with the help of Maxwell's equations, one can derive the wave vector of a electromagnetic wave mode that propagates along the metal-dielectric interface \cite{Chubchev}
\begin{equation}\label{dysp_bulk}
\kappa(\omega)=\kappa'+i\kappa''=\frac{\omega}{c}\sqrt{\frac{\epsilon_1\epsilon_2}{\epsilon_1+\epsilon_2}},
\end{equation}
where $\kappa'$ and $\kappa''$ are the real and imaginary parts of the wave vector component, respectively and $c$ is the speed of light in vacuum. For real values of $\epsilon_1$ and $\epsilon_2$, one obtains real wave vector when $\epsilon_1\epsilon_2<0$ and $\epsilon_1+\epsilon_2<0$ and in such a case propagating SPPs can be excited.

\begin{figure}[ht!]
a)\includegraphics[width=.8\linewidth]{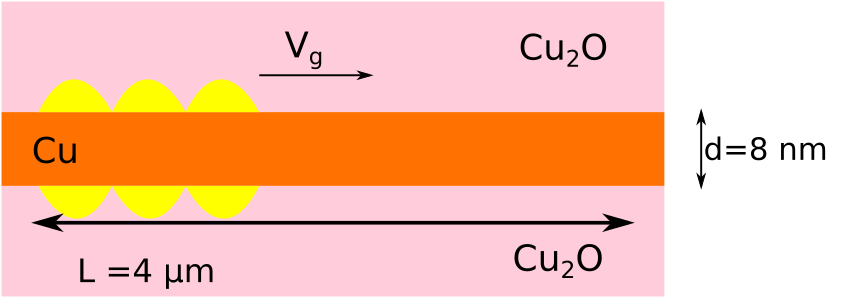}
b)\includegraphics[width=.8\linewidth]{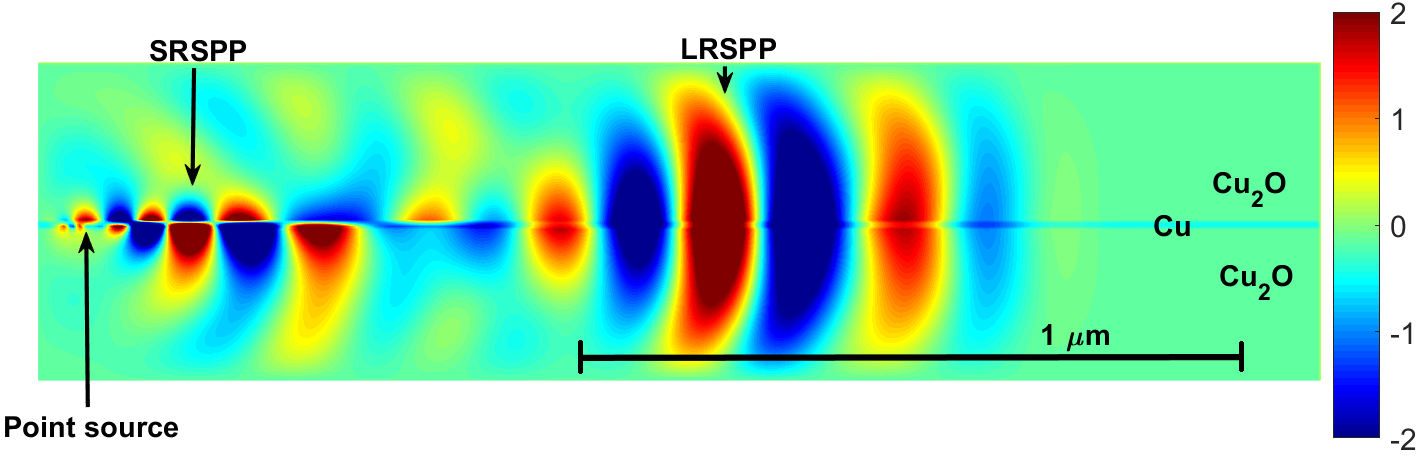}
\caption{a) Schematic representation of the system considered in the paper; surface plasmon (yellow) propagates along the metal surface at some group velocity $V_g$; b) Spatial distribution of normalized electric field (color) obtained in FDTD simulation.}\label{fig:schemat}
\end{figure}

In this paper, we will consider a metal layer of finite thickness $d$. In such a case, one obtains modified boundary conditions which result in two solutions \cite{Raether1988,My_OL}
\begin{eqnarray}\label{dyspersyjna}
\frac{\kappa_1\tanh(\kappa_1\frac{d}{2})}{\epsilon_1} = \frac{-\kappa_{2}}{\epsilon_{2}},\nonumber\\
\frac{\kappa_1\coth(\kappa_1\frac{d}{2})}{\epsilon_1} = \frac{-\kappa_{2}}{\epsilon_{2}},
\end{eqnarray}
where $\kappa_{1,2} = \sqrt{k^2 - \epsilon_{1,2}\frac{\omega^2}{c^2}}$. In the limit of $d \rightarrow \infty$, Eqs. (\ref{dyspersyjna}) reduce to the form given by Eq. (\ref{dysp_bulk}). The two above solutions are the so-called short-range and long-range plasmons (SRSPP, LRSPP accordingly). They are characterized by a small or large group velocity
\begin{equation}\label{eq:vg}
V_g = \frac{\partial\omega}{\partial \kappa'} = \frac{c}{n_{eff}+\omega\frac{\partial n_{eff}}{\partial \omega}},
\end{equation}
where $n_{eff}=\sqrt{\epsilon}$ is the effective refraction index. The crucial property of long-range plasmons is reduced absorption and elevated group velocity, which allows them to propagate over distances on the order of tens to hundreds of $\mu$m \cite{Konopsky2006,Yi2007,Park2009,Berini2009}. One can take an advantage of this to produce relatively long-living plasmons in copper. As mentioned before, in some cases the ohmic losses in Cu-based systems can approach the values typical to silver and gold nanostructures, provided that the detrimental effects of oxide layer is mitigated. Here, we propose a system where the Cu$_2$O is an inherent part of the structure necessary for its operation. Particularly, we consider a thin layer of Cu surrounded by semi-infinite (much thicker than wavelength) layers of Cu$_2$O. The system is shown on the Fig. \ref{fig:schemat} a). An external light source (point source of radiation in Finite Difference Time Domain (FDTD) calculations, described in Appendix A) generates both long- and short range excitons; they can be seen on the Fig. \ref{fig:schemat} b), where the field distribution is symmetric or antisymmetric for LRSPP and SRSPP, accordingly.

\begin{figure}[ht!]
\includegraphics[width=.8\linewidth]{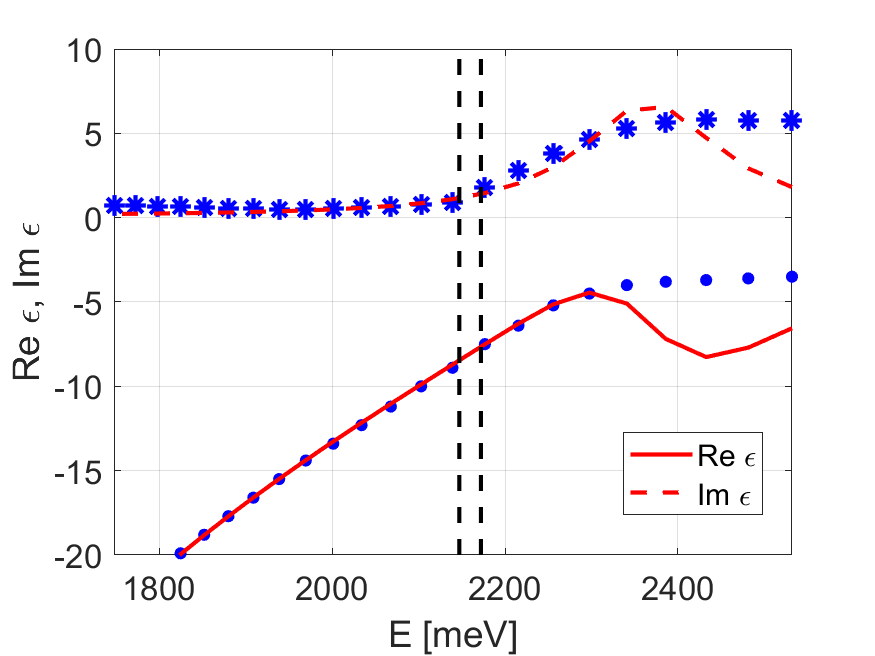}
\caption{Permittivity data from \cite{Hollstein} (dots) and the fitted model (lines). Dashed black lines mark the part of the spectrum occupied by Rydberg excitons in Cu$_2$O, indicating energy of 2P exciton and gap energy.}\label{fig:model}
\end{figure}

In order to derive the optical properties of the proposed system, one needs a suitable model of the materials. In the considered spectral range of $2-2.2$ eV, the permittivity of the oxide is very close to a constant $\epsilon_2 = 7.5$, with the  additional impact of excitons as outlined in the next section. For the properties of copper, we refer to \cite{Hollstein}. The real and imaginary part of permittivity and the fitted model used in FDTD calculations are shown on the Fig. \ref{fig:model}. 
The model is fitted to provide a good match in the spectral region of Rydberg excitons, marked by dashed lines; specifically, the maximum difference between experimentally measured susceptibility and model is less than 2$\%$ in a 100 meV range around the 2P exciton energy, which is considerably more than the spectral width of simulated plasmons. Two main features of the susceptibility relation $\epsilon_1(\omega)$ are apparent; there is a considerable increase of absorption (proportional to the imaginary part of permittivity), which is the result of inter-band transitions in Cu \cite{Hollstein}. This increase starts just above the gap energy of Cu$_2$O, so that the excitons, which have energies smaller than the band gap, still remaining in the low absorption part of the spectrum. Furthermore, the real part of permittivity is negative and in fact in the spectral region of excitons, it provides a close match to the oxide permittivity, e. g. $\epsilon_1+\epsilon_2 \approx 0$. This is a crucial condition for excitation of strong plasmonic resonance; these findings are consistent with \cite{Takagi17}, where a strong plasmonic resonance in the region of 2 eV has been demonstrated in a thin copper film.  

\section{Tunable long range plasmons}
\begin{figure}[ht!]
\includegraphics[width=.9\linewidth]{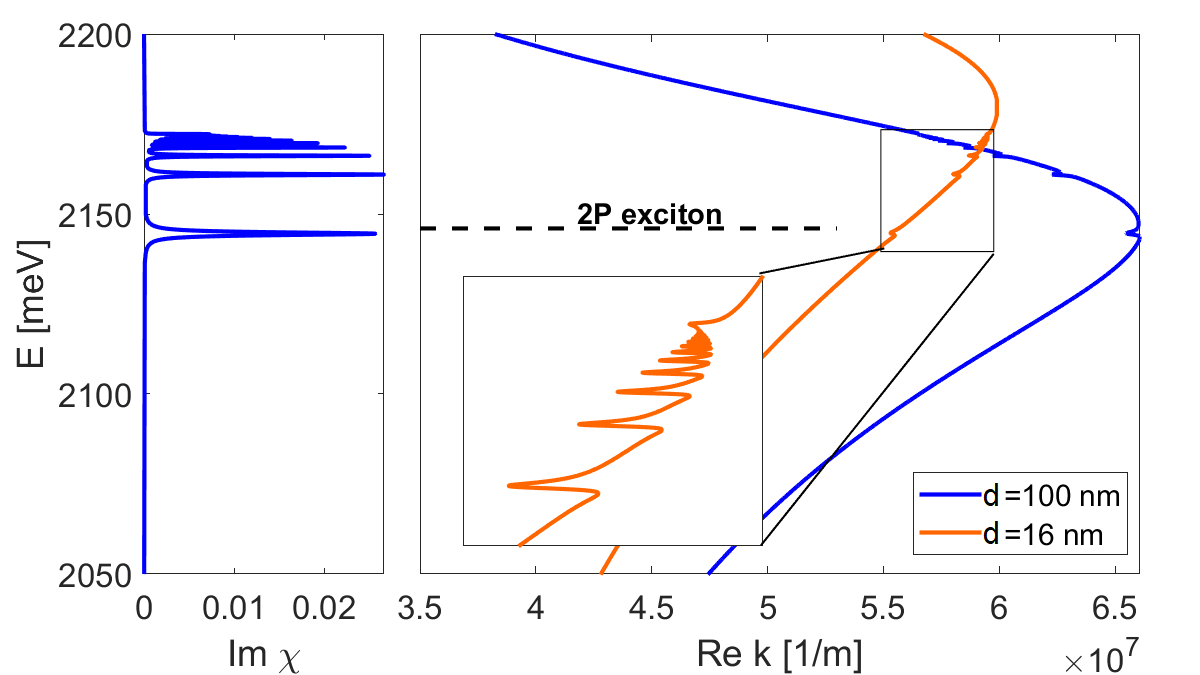}
\caption{Dispersion relation of surface plasmon calculated for a case of thick and thin layer of Cu.}\label{fig:dysp}
\end{figure}

Let's consider a long-range plasmon propagating along a thin layer of Cu surrounded by Cu$_2$O. By using the above-discussed model of metal permittivity $\epsilon_1$ and the oxide permittivity $\epsilon_2$ containing a numerically calculated excitonic spectrum \cite{PRB93}, we can calculate the exciton wave vector from Eq (\ref{dyspersyjna}). The results are shown on the Fig. \ref{fig:dysp}. 
The left panel shows the Cu$_2$O susceptibility $\chi(\omega)=\epsilon_2(\omega)-7.5$ exhibiting the characteristic spectrum of Rydberg excitons with multiple states corresponding to principal quantum number $n=1,2...20$. These resonances have a noticeable impact on the dispersion curve $\kappa(\omega)$ (right panel, inset). While the absolute value of the changes of $\epsilon_2$ caused by excitons is rather small, these changes occur in a very narrow spectral range of excitonic resonances. This means that the group velocity of plasmon, which is connected with the slope of the function $\kappa(\omega)$ according to Eq. (\ref{eq:vg}), is significantly affected by the excitons. This is a key component of our proposal of tunable plasmons. Due to the change of the plasmon group velocity, the time needed to propagate some certain distance is changed. This, in turn, affects the amount of plasmon energy absorbed by the metal due to the ohmic losses. As a result, we can significantly change the transmission coefficient of the system despite the fact that direct absorption by Cu$_2$O is negligible when compared to Cu.
Another crucial component of our proposal is the use of a thin Cu layer. One can see that the dispersion relation on the Fig. \ref{fig:dysp} calculated for a thick layer changes from a normal dispersion below $2.14$ eV to anomalous dispersion above (e.g. the value of $\kappa$ is decreasing with energy). However, by using a sufficiently thin layer, one can increase the energy where transition to anomalous dispersion occurs; the same technique has been used in \cite{my_UV} to facilitate propagation of UV plasmons. In the case here, it allows for excitation of plasmons in the $2.1-2.2$ eV range, fitted to the spectral region of Rydberg excitons. Specifically, Fig. \ref{fig:dysp} inset shows the dispersion relation of such plasmons. One can see that the general slope of the function $\omega(\kappa)$ is steeper than in the thick layer case, indicating increased group velocity of LRSPPs. At the same time, the slope of $\omega(\kappa)$ near excitonic resonances is considerably smaller, corresponding to greatly reduced group velocity. As discussed in \cite{Stolz2020}, group velocities on the order of 10$^3$ m/s are possible for a light pulse tuned to $n>10$ exciton propagating in Cu$_2$O. In the case of low $n$ excitonic states and plasmons considered here, the slowdown is less considerable; since the plasmon dispersion relation depends on the permittivity of both dielectric and metal, the effect of excitonic resonances is reduced. Furthermore, the spectrum of the plasmon is much wider than the width of excitonic resonance, so that only selected spectral components of SPP are affected by the exciton. This is discussed further below. Finally, from practical point of view, apart from potential irregularity of oxidation film (Cu$_2$O), the close separation of higher excitonic resonances prevents them from being selectively excited by a relatively short-living, wide spectrum plasmon.

\begin{figure}[ht!]
\includegraphics[width=.8\linewidth]{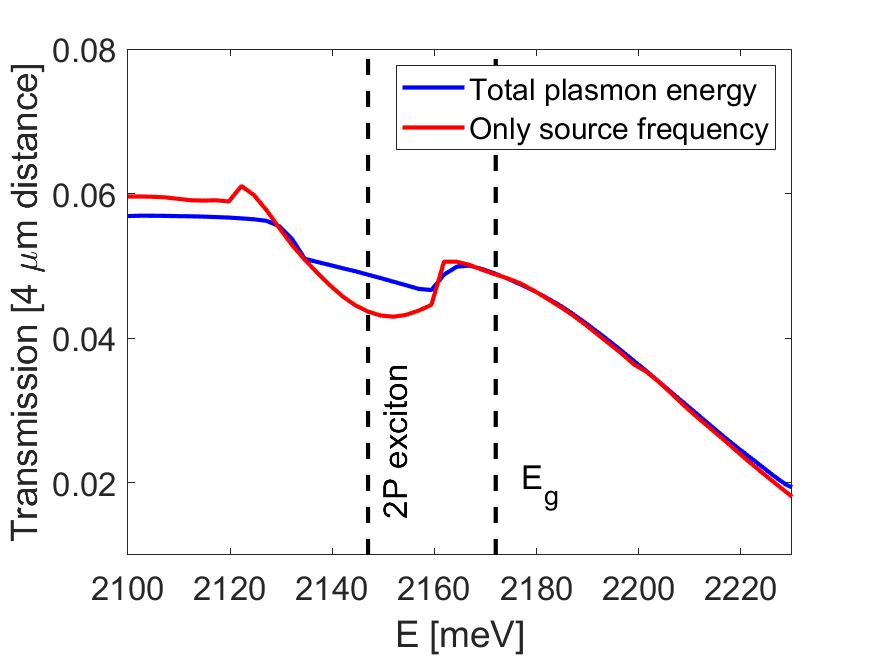}
\caption{Transmission coefficient of long range plasmons on a thin Cu layer.}\label{fig:transmission}
\end{figure}

The FDTD results, where the transmission coefficient of LRSPP propagating through a distance of $L=4$ $\mu$m, on a $d=8$ nm thick Cu layer was calculated, is shown on the Fig. \ref{fig:transmission}.
Notably, transmission coefficient on the order of 5$\%$ is obtained, which is considerably bigger than can be achieved with bulk plasmons, even with noble metals \cite{Berini2009}. The two transmission coefficients shown on the Fig. \ref{fig:transmission} correspond to the entire plasmon and its central frequency. One can see a noticeable drop of transmission in the region of the strongest 2P exciton. This minimum is more pronounced when only the central frequency is considered. This is caused by the fact that the spectrum of propagating plasmon is considerably wider that the excitonic resonance. Specifically, in this particular case the plasmon lifetime is on the order of $10^{-13}$ s, corresponding to the spectral width $\Delta E \sim 30$ meV. This value is comparable to the distance between $n=2$ and $n=3$ excitonic states (14 meV) and approximately 6 times larger than the widest resonance of $n=2$ exciton (5 meV) \cite{Kazimierczuk}. Correspondingly, the exciton lifetime is several times longer than plasmonic lifetime, being equal to $\tau \sim 1$ ps for $n=2$. Finally, while the low-n excitons are considerably smaller than the space occupied by propagating plasmon (2P exciton radius is approx. $r_E=1.5$ nm, blockade radius $r_B=19$ nm), they can easily exceed the plasmon size for larger n. This particular set of size and lifetime proportions has several important consequences:
\begin{itemize}
\item The local reduction of the group velocity is applied only to some wave modes close to the central frequency of excitonic resonance and not entire plasmon spectrum; this means that the effectiveness of slowdown of a surface plasmon as a whole is reduced and in general, a single group velocity cannot be attributed to the plasmon (e.g. there is considerable group velocity dispersion).
\item By fine-tuning the frequency of the light exciting the plasmon, one can effectively aim at $n=2$ and $n=3$ excitonic states; higher states are located too closely to each others to be targeted selectively in such a system. However, this restriction does not apply when a second, non-plasmon field is used to excite some specific state, which is then used in exciton-plasmon interaction. 
\item The locally enhanced absorption of specific modes produces a narrow dip in the plasmon spectrum, similar to the so-called spectral hole-burning \cite{Wahid}.
\end{itemize}

\begin{figure}[ht!]
\includegraphics[width=.9\linewidth]{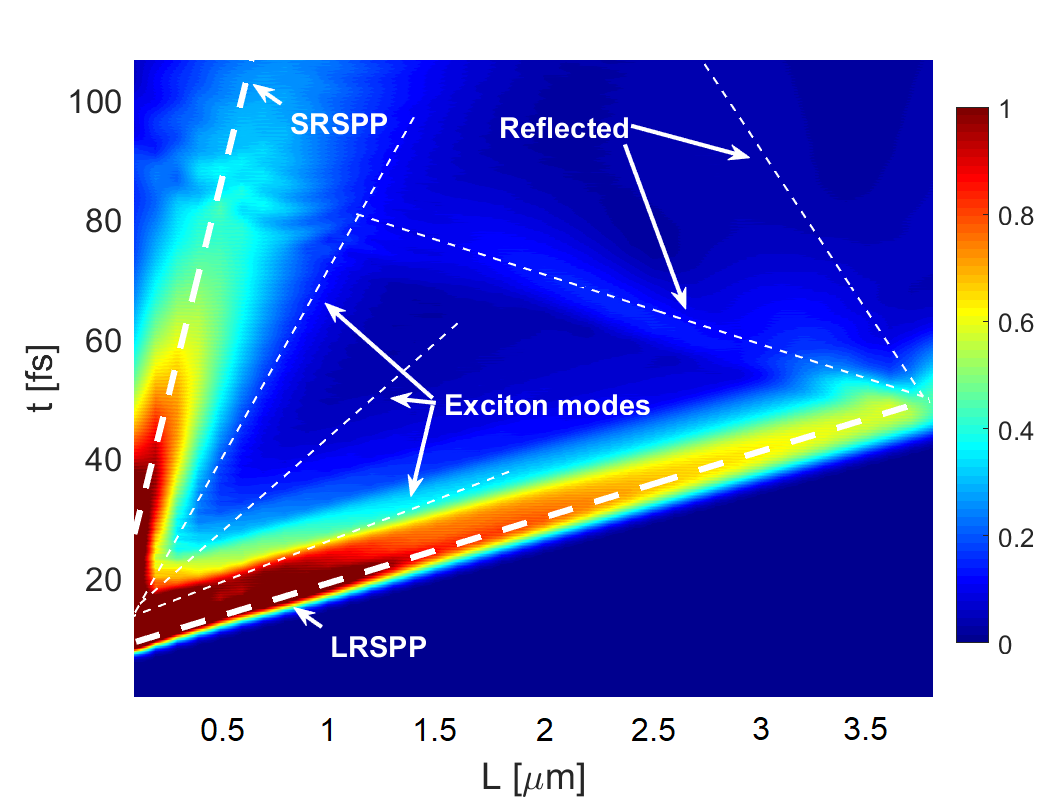}
\caption{Normalized field amplitude (color) of excited plasmonic modes as a function of time and position along the metal layer.}\label{fig:propagation}
\end{figure}

To confirm the above findings and further investigate the plasmon dynamics and its interaction with excitons, one can analyse the group velocity of excited plasmonic modes by considering the electric field amplitude as a function of time and position, as shown on the Fig.\ref{fig:propagation}. 
The figure is dominated by two strong lines representing LRSPP and SRSPP modes; their slope differs considerably, indicating group velocities on the order of $0.3$ c and $0.04$ c, accordingly. One can also see that the long-range plasmon maintains non-negligible amplitude after reaching the end of computation domain at L=4 $\mu m$, and then it is reflected. At the point of reflection, boths long-range and short-range modes are created (more horizontal and vertical lines, accordingly). Finally, one can notice weaker traces of several modes with group velocities that differ considerably from the LRSPP. These modes  correspond to the spectral range of 2P exciton, specifically in the regions of increased normal dispersion around the excitonic resonance. As mentioned before, due to the reduction of group velocity, these modes are more strongly absorbed by the metal. Moreover, since the group velocity change caused by excitonic resonance occurs in a very narrow part of the spectrum, being only a fraction of the plasmon spectrum, the slow modes carry relatively little energy and thus are barely visible on the Fig. \ref{fig:propagation}. A somewhat explicit picture can be obtained by considering the image of the plasmon in frequency domain. 
\begin{figure}[ht!]
\includegraphics[width=.9\linewidth]{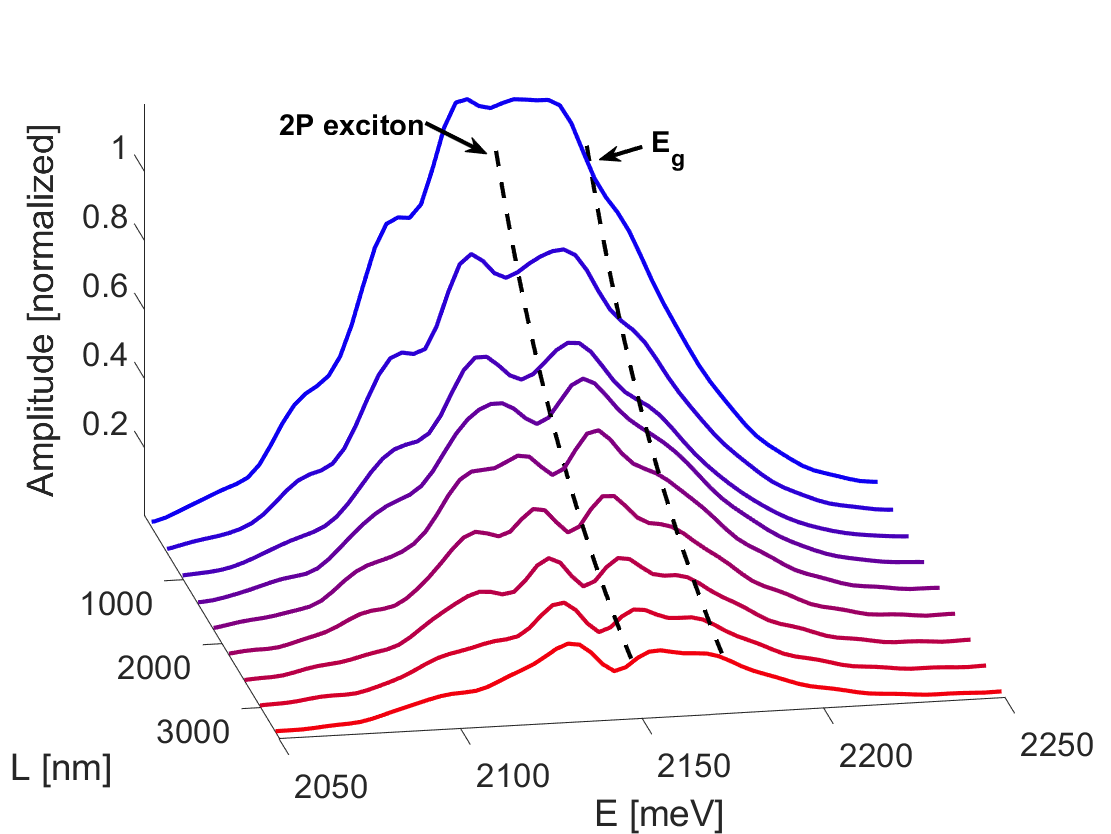}
\caption{Frequency spectra of the propagating LRSPP calculated at various propagation lengths.}\label{fig:pspectrum}
\end{figure}
The spectrum of LRSPP is shown on the Fig. \ref{fig:pspectrum}. One can see how the plasmon spectrum evolves with the propagation length; initially, it is approximately a Gaussian with the width on the order of 100 meV. However, as the plasmon propagates along the metal surface, a noticeable dip is formed at the energy of 2147 meV, which corresponds to 2P exciton resonance. Another, weaker local minimum is present above the band gap energy $E_g=$2172 meV. This means that increased absorption of Cu$_2$O in the region above the band gap is a relatively minor contribution to absorption as compared to the slowdown of $V_g$ and subsequent absorption by Cu. After the propagation distance of $L=3\mu $m, the local minimum in the spectrum is quite substantial, with approximately 30$\%$ lower amplitude. This is consistent with the results in the Fig. \ref{fig:transmission}, where a local reduction of transmission coefficient from $6\%$ to $4\%$ is shown. It should be stressed that the LRSPP is not an ideal Gaussian but a collection of modes that propagate at various group velocities that can differ considerably; this complicates calculation of the spectrum and is the cause of other local minima visible on the Fig. \ref{fig:pspectrum}. However, in contrast to the minimum corresponding to the 2P exciton, these apparent minima do not appear consistently in each spectrum, at the same energy. Finally, from the Fig. \ref{fig:schemat} b) one can see that the electric field of plasmons extends into Cu$_2$O up to a distance of $\sim 300$ nm from the metal surface; therefore, the plasmon field can easily reach excitons that are not strongly affected by the immediate vicinity of metal-oxide interface and the use of unperturbed energy of 2P exciton state $E_{2P} = 2147$ meV in calculations is justified.

As mentioned before, the strength of the optical response of Cu$_2$O is directly proportional to the exciton density. This, in turn, depends on the energy provided to generate the excitons and is limited by the Rydberg blockade. As a result, one has several options for dynamically controlling the system:
\begin{itemize}
\item When the propagating plasmon's frequency is tuned to the excitonic resonance, its energy can be used to create new excitons. In such a case, the response of the system (transmission coefficient) will be highly dependent on the power of the plasmon.
\item One can use external optical field (control field) to create a given density of excitons, which then affects the propagation of weaker probe field (plasmon).
\item Due to the narrow spectrum of excitonic resonances, only a slight detuning of plasmon frequency is needed to considerably affect its transmission.
\item By exciting two plasmons in a short period, one can potentially create conditions where the excitons generated by the first plasmon affect the propagation of the second plasmon, allowing for exciton-mediated plasmon-plasmon interactions.
\item In a similar manner, one can arrange interaction between plasmons propagating along two closely spaced metal layers.
\end{itemize}

\begin{figure}[ht!]
\includegraphics[width=.9\linewidth]{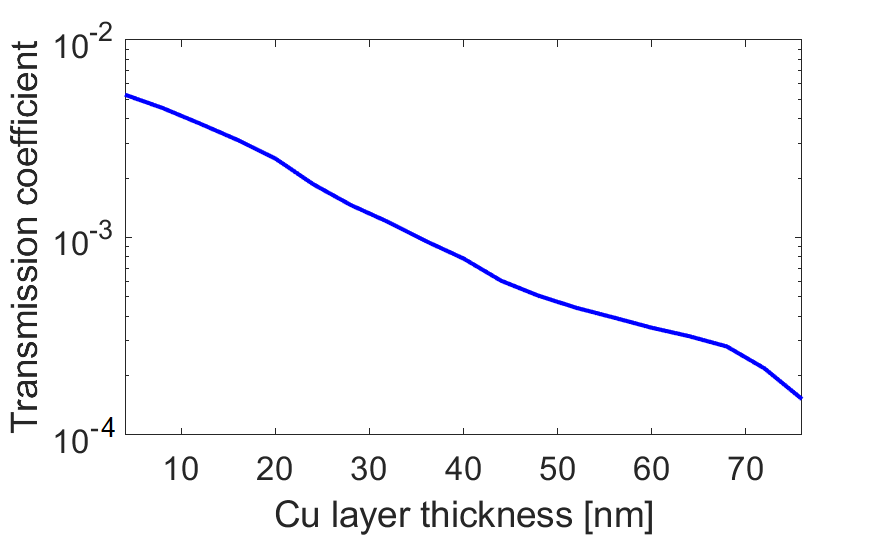}
\caption{Transmission coefficient over the distance $L=4$ $\mu$m as a function of metal layer thickness $d$.}\label{fig:tthick}
\end{figure}

From the point of view of practical applications, it is important to consider the maximum thickness of Cu layer that allows for propagation of long-range plasmons. As mentioned in \cite{Berini2009}, the thickness $d$ should be considerably smaller than the plasmon wavelength. On the fig. \ref{fig:schemat} b), one can estimate that this wavelength is about 200 nm (vacuum wavelength of incident light is $\lambda=577$ nm, which indicates that effective refraction index is $n_{eff} \approx 2.9$; this is slightly larger than the index of Cu$_2$O $n_{2} = 2.73$). Therefore, one can assume that the metal layer thickness must be $d << 200$ nm. This is the case on the Fig. \ref{fig:tthick}; the transmission coefficient decreases exponentially with the layer thickness,with a value of $T\sim 4 \%$ for $d$=10 nm and $T \sim 0.1 \%$ for $d=$35 nm.
\section{FDTD method}
One of the most popular tools for numerical analysis of SPP propagation is the Finite-Difference Time-Domain (FDTD) method \cite{Yee}. It is based directly on Maxwell's equations, which provides a high flexibility and accuracy in modelling the propagation of electromagnetic waves. In this manuscript, we focus on the two-dimensional variant of the method; the computational domain is a cross-section of the system placed on $xy$ plane. The structure is assumed to be symmetric and semi-infinite (e.g. much larger than light wavelength) in $z$ direction. The domain is divided into rectangular grid of discrete cells with the size $\Delta x = 4$ nm. In every cell, the values of the components of the fields $\vec{E}(x,y,t)$, $\vec{H}(x,y,t)$ are stored. By defining a discrete time step $\Delta t$, one can use the Maxwell's equations to derive evolution equations that use the field values $\vec{E}(x,y,t)$, $\vec{H}(x,y,t)$ to calculate the next ones $\vec{E}(x,y,t+\Delta t)$, $\vec{H}(x,y,t+\Delta t)$. In the two-dimensional case, without the loss of generality one can assume that the fields have only three non-zero components $\vec{E}=[E_x,E_y,0]$, $\vec{H}=[0,0,H_z]$.
The optical response of materials in the simulation is described by the polarization vector $\vec{P}(x,y,t)$. Specifically, one can use the so-called Drude-Lorentz model that specifies the connection between electric field in polarization in the frequency domain
\begin{eqnarray}\label{drude}
\vec{P}(\omega)&=&\epsilon(\omega)\vec{E}(\omega),\nonumber\\
\epsilon(\omega)&=&\epsilon_\infty+\sum_{j=1}^{n}\frac{\omega_{pj}^2}{\omega_{0j}^2 - \omega^2 - i\gamma_j\omega}
\end{eqnarray} 
with a set of $n$ oscillators with the so-called plasma frequency $\omega_{pj}$, resonance frequency $\omega_{0j}$ and dissipation constant $\gamma_j$. In the case of Cu, one can use two oscillators; the first one, with $\omega_{p1}=0$ describes the baseline Drude model of free electrons in metal, while the second oscillator provides the correction responsible for the increase of absorption due to the inter-band transitions (the local maximum at $\omega=0.047$ on the Fig. \ref{fig:model}). For Cu$_2$O, one has $\epsilon_\infty=7.5$ and the series of oscillators corresponding to excitonic resonances, with appropriate oscillator strengths and damping constants. The medium model has been verified by comparing the numerically obtained absorption spectrum of Cu$_2$O with the results in \cite{Kazimierczuk}. In the performed calculations, the plasmon center frequency is tuned to $n=2$ exciton and thus an inclusion of two excitons $n=2$ and $n=3$ provides sufficient accuracy of results.
The above medium frequency response is integrated into time domain simulation with the use of ADE (Axillary Differential Equations) approach \cite{Alsunaidi}. In this method, one can calculate the polarization as a function of time with the following differential equation
\begin{equation}\label{polaryzacje}
\ddot{P}+\gamma_j\dot{P}+\omega^2_{0j} P=\frac{\omega^2_{pj}}{\epsilon_\infty} E
\end{equation}
for every oscillator $j$. The final set of equations based on Eq. \ref{polaryzacje} and Maxwell's equation is as follows
\begin{eqnarray}\label{FDTD_finalne}
&&\frac{\partial E_y(x,y,t)}{\partial x}-\frac{\partial E_x(x,y,t)}{\partial y}=-\mu_0\frac{\partial H_z(x,y,t)}{\partial t},\\
&&\frac{\partial H_z(x,y,t)}{\partial y}=j_{x}(x,y,t)+\epsilon_0\frac{\partial E_x(x,y,t)}{\partial t} -\frac{\partial P_x(x,y,t)}{\partial t},\nonumber\\
&&-\frac{\partial H_z(x,y,t)}{\partial x}=j_{y}(x,y,t)+\epsilon_0\frac{\partial E_y(x,y,t)}{\partial t} +\frac{\partial P_y(x,y,t)}{\partial t}.\nonumber\\\nonumber
\end{eqnarray}
Where the polarization has two components $P_x$, $P_y$ which depend on $E_x$, $E_y$ according to Eq. (\ref{polaryzacje}). The $j_x$ and $j_y$ are current densities in the corresponding directions and $\epsilon_0$, $\mu_0$ are vacuum permittivity and permeability. By rearranging the terms in Eq. (\ref{FDTD_finalne}), one can express the future values of the fields $E_x(x,y,t+1)$, $E_y(x,y,t+1)$, $H_z(x,y,t+1)$ as a function of the current values. Such relations allow for advancing the state of the system in discrete time steps by calculating the new field values basing on the current ones. 

\section{Conclusions}
The discovery of Rydberg excitons in Cu$_2$O has opened many new perspectives in semiconductor science, but also posed many theoretical problems and challenges. While the optical properties of REs, especially in bulk media, are already well understood, the ongoing research on Rydberg excitons is now entering the phase where the first experiments involving confined systems are performed. Thus, it is of high importance to provide the theoretical framework for understanding the properties of these highly complex media. Metallic nanostructures and exciton-plasmon interaction are one of the promising directions of study.
Due to the inherent losses in Cu, regular plasmon modes alone are not suitable as carriers of information over the long distances. However, as we have demonstrated, the so-called long-range plasmons can achieve propagation length on the order of micrometres. At this scale, it is possible to couple them with excitons in Cu$_2$O, allowing for many interesting options to dynamically control the plasmon propagation. It is shown that the group velocity of propagating plasmon can be significantly affected by the presence of excitons, which affects the system transmission. The presented results may pave the way to compact, copper-based tunable devices.

\end{document}